\documentstyle[prl,aps,epsfig,twocolumn]{revtex}
\def\break#1{\pagebreak \vspace*{#1}}

\begin{document}

\setlength{\parindent}{0.4cm}

\title{Controlling spatiotemporal chaos in excitable media by local biphasic 
stimulation}

\author{Johannes Breuer{\footnote{Also at: Institute of Theoretical Physics,
Technical University-Berlin, D-10623, Berlin, Germany}} and Sitabhra
Sinha{\footnote{Electronic address: sitabhra@imsc.res.in}} }

\address{The Institute of Mathematical Sciences, C. I. T. Campus,
Taramani, Chennai - 600 113, India.}

\maketitle

\widetext
\begin{abstract}
Controlling spatiotemporal chaos in excitable media by applying low-amplitude
perturbations {\em locally} is of immediate applicability, e.g.,  in treating
ventricular fibrillation, a fatal disturbance in the normal rhythmic
functioning of the heart. We look at a mechanism of control by the local
application of
a series of biphasic pulses, i.e., involving both positive 
and negative stimulation. 
This results in faster recovery of the medium, making
it possible to overdrive the chaos
by generating waves with frequency higher than that possible with
only positive pulses.
This provides the simplest and most general understanding of the
effectiveness of biphasic stimulation in controlling fibrillation and
allows designing optimal waveshapes for controlling spatiotemporal chaos.

\vspace{0.25cm}
\end{abstract}

\vspace{0.63cm}
\narrowtext
A characteristic feature of excitable media is the formation of
spiral waves and their subsequent breakup into spatiotemporal chaos.
Examples include catalysis of CO on Pt(110) surface \cite{Hil95},
cAMP waves during slime mold aggregation \cite{Vas94}, etc.
Another example of obvious importance is the propagation of 
waves of electrical excitation along the
heart wall, initiating the muscular contractions that enable
the heart to pump blood. 
In fact, spiral turbulence has been identified by several
investigators as the underlying cause of certain arrhythmias, i.e.,
abnormal cardiac rhythms, including ventricular 
fibrillation (VF) \cite{Gra98}, a potentially fatal condition in which
different regions of the heart are no longer activated coherently.
Current methods of defibrillatory 
treatment involve applying large electrical
shocks to the entire heart in an attempt to drive it to the normal state.
However, this is not only painful but also dangerous, as the resulting 
damage to heart
tissue can form scars that act as substrates for future cardiac
arrhythmias. Devising a 
low-amplitude control mechanism for spatiotemporal chaos in excitable
media is therefore not only an exciting theoretical challenge but of
potential significance for the treatment of VF. In this paper, we propose
using low-amplitude biphasic pacing, i.e., applying a sequence of 
alternating positive and negative
pulses locally, as a robust control method for such chaos.

Most of the methods proposed for controlling spatiotemporal chaos in 
excitable media involve applying perturbations either globally or
over a spatially extended system of control points covering
a significant proportion of the entire system \cite{Osi99,Rap99,Sin01,Wan03}. 
However, in most real situations this may not be a feasible option. 
Further, in the specific context of controlling VF, a local control scheme
has the advantage that it can be readily implemented with existing hardware
of the Implantable Cardioverter-Defibrillator (ICD). This is a device
implanted into patients at high risk from VF that monitors the
heart rhythm and applies electrical treatment, when necessary, through
electrodes
placed on the heart wall. 
\break{3.1cm}
A low-energy control method 
involving ICDs should therefore aim towards achieving control of
spatiotemporal chaos by applying small perturbations from a
few local sources.

For most of the simulations reported in this paper 
we have used the modified Fitzhugh-Nagumo
equations proposed by Panfilov as a model for ventricular activation 
\cite{Pan93}.
For simplicity we assume an isotropic medium; in this case the model
is defined by the two equations governing the excitability $e\/$ and recovery
$g\/$ variables,
\begin{equation}
\begin{array}{lll}
{\partial e}/{\partial t} & = & {\nabla}^2 e - f(e) - g,\\
{\partial g}/{\partial t} & = & {\epsilon}(e,g) (ke - g).
\end{array}
\label{panfeq}
\end{equation}
The function $f(e)\/$, which specifies fast processes (e.g.,
the initiation of 
excitation)
is piecewise linear:
$f(e)= C_1 e$, for $e<e_1$, $f(e) = -C_2 e + a$,
for $e_1 \leq e \leq e_2$, and $f(e) = C_3 (e - 1)$, for $e > e_2$.
The function $\epsilon (e,g)$, which determines the dynamics of the
recovery variable, is $\epsilon (e,g) = \epsilon_1$ for
$e < e_2$, $\epsilon (e,g) = \epsilon_2$ for $e > e_2$, and
$\epsilon (e,g) = \epsilon_3$ for $e < e_1$ and $g < g_1$. 
We use the physically appropriate parameter values given
in Ref.\cite{Sin01}.

We solve model (1) by using a forward-Euler integration scheme. The system
is discretized on a two-dimensional grid of points with spacing
$\delta x = 0.5\/$ dimensionless units. The standard five-point 
difference stencil is used for the 2-D Laplacian. The spatial grid consists of
$L \times L$ points;
in our studies we have used values of $L\/$ upto 500. For the 1-D 
simulations we use a 3-point
difference stencil for the Laplacian, with the spatial lattice
consisting of $L$ points.
The time step is $\delta t = 0.01$ dimensionless units.
We did not observe any qualitative change in the results on decreasing the
space and time steps by a factor of 2. 
On the edges of the
simulation region we use no-flux (Neumann) boundary conditions.

To achieve control of spatiotemporal chaos, we locally apply
a periodic perturbation, $A F ( 2 \pi f t)$, of amplitude $A$ and frequency
$f$. $F$ can represent any periodic function, e.g., a series of pulses
having a fixed amplitude and duration, applied at periodic intervals defined by 
the stimulation frequency $f$. 
The control mechanism can be understood as a process of overdriving the
chaos by a source of periodic excitation having a significantly higher 
frequency.
As noted in Refs. \cite{Lee97,Xie99}, in a competition between two sources of
high frequency stimulation, the outcome is independent
of the nature of the wave generation at either source, and is decided solely
on the basis of their relative frequencies. This follows from
the property of an excitable medium that waves annihilate when they
collide with each other \cite{Kri83}. The lower frequency source is
eventually entrained by the other source and 
will no longer generate waves
when the higher frequency source is withdrawn. 
Although we cannot speak of a single frequency source 
in the case of chaos, the
relevant timescale is that of spiral waves which is limited by the
refractory period of the medium, $\tau_{ref}$, the time interval during
which an excited cell cannot be stimulated until it has recovered its
resting state properties. To achieve control, one
must use a periodic source with frequency $f > \tau_{ref}^{-1}$. This is
almost impossible 
with purely excitatory stimuli as reported
in Ref.\cite{Sin01}; the effect of locally applying such perturbations is 
essentially limited by refractoriness to the immediate neighborhood 
of the stimulation point.

A simple argument shows why a negative rectangular pulse decreases the
refractory period for the Panfilov model in the absence of the diffusion term. 
The stimulation vertically displaces
the $e$-nullcline of Eq.~(\ref{panfeq})
and therefore, the maximum value of $g$ that can be attained
is reduced. Consequently, the system will recover 
faster from the refractory state. 
To illustrate this, let us assume that the stimulation is applied 
when $e > e_2$.
Then, the dynamics reduces to $\dot e = -C_3 (e - 1) - g, 
\dot g = \epsilon_2 (k e - g)$. In this region of the
($e, g$)-plane, for sufficiently high $g$, the trajectory will be along
the $e$-nullcline, i.e., $\dot e \simeq 0$.
If a pulse stimulation of amplitude $A$ is initiated 
at $t = 0$ (say), when $e = e (0), 
g = g (0)$, 
at a subsequent time $t$, $e ( t ) = 1 + {\frac{A - g ( t )}{C_3}}$, 
and $g ( t ) = {\frac{a}{b}} - [{\frac{a}{b}}- g(0)] exp(- b t)$, where,
$a = \epsilon_2 k (1+ {\frac{A}{C_3}}), 
b = \epsilon_2 [1 + (k/C_3)]$. 
The negative stimulation has to be kept on till the
system crosses into the region where $\dot e < 0$, after which
no further increase of $g$ can occur, as dictated by the dynamics
of Eq. (\ref{panfeq}).
Now, the time required by the system to enter this region
is 
${\frac{1}{b}} {\rm ln} {\frac{a/b - g (0)}{a/b - \phi}}$, 
where $\phi = C_3 ( 1 - e_2 ) + A$.
Therefore, this time is reduced when $A < 0$ and contributes to the decrease
of the refractory period.
Note that, the above discussion also
indicates that a rectangular pulse will be more effective than a gradually
increasing waveform, e.g., a sinusoidal wave (as used in \cite{Zha03}), 
provided the energy of stimulation is same in both cases,
as the former allows a much smaller maximum value of $g$. 
Therefore, phase
plane analysis of the response to negative stimulation allows
us to design waveshapes for maximum efficiency in controlling spiral
turbulence.

To understand how negative stimulation affects the response behavior
of the spatially extended system, we first look at a one-dimensional
system. Fig.~1  shows the relation between the stimulation frequency $f$
and effective frequency $f_{eff}$, measured by applying a series of pulses 
at one site and then recording the number of pulses that reach another site
located at a distance without being blocked by any refractory region.
Depending on the relative value of $f$ and $\tau_{ref}$, we observe
instances of $n : m$ response, i.e., $m$ responses evoked by $n$ stimuli.
From the resulting effective frequencies $f_{eff}$, we can see that for
purely excitatory stimulation, the relative refractory
period can be reduced by increasing the amplitude $A$. 
However, this reduction is far more pronounced when 
a negative stimulation is applied between every pair of positive pulses.
The inset in Fig.~1 shows that there is an optimal time interval
between applying the positive and negative pulses that 
decreases the refractory period by as much as $50 \%$. 
The highest effective frequencies correspond to 
a stimulation frequency in the range $0.1-0.25$, agreeing with the optimal
time period of 2-5 time units between positive and negative stimulation.

A response diagram similar to the one-dimensional case is also seen
for stimulation in a two-dimensional medium (Fig.~2).
A small region consisting of $n \times n$ points
at the center of the simulation domain is chosen as the stimulation point.
For the simulations reported here $n = 6$; for a smaller $n$, one requires
a perturbation of larger amplitude to achieve a similar response.
To understand control in two dimensions, we find out the characteristic
time scale of spatiotemporal chaotic activity by obtaining its power
spectral density (Fig.~2, inset). We observe a peak at a frequency 
$f_c \simeq 0.0427$. As seen in Fig.~2, there are ranges
of stimulation frequencies that give rise to effective frequencies
higher than this value. As a result, the periodic waves emerging
from the stimulation point will gradually impose control over the 
regions exhibiting chaos.
If $f$ is only slightly higher than $f_c$
control takes very long; if it is too high
the waves suffer conduction block at inhomogeneities produced by 
chaotic activity that reduces the effective frequency, and control fails. 
Note that, at lower frequencies the range of stimulation frequencies 
for which $f_{eff} > f_c$, is smaller than 
at higher frequencies. 
We also compare the performance of sinusoidal waves
with rectangular pulses, adjusting the amplitudes so that they
have the same energy. The former is much less effective than the latter
at lower stimulation frequencies, which is the preferred operating region  
for the control method.

The effectiveness of overdrive
control is limited by the size of the system sought to be controlled.
As shown in Fig.~3, away from the control site, the generated waves
are blocked by refractory regions, with the probability of block
increasing as a function of distance from the site of stimulation\cite{note1}.
To see whether the control method is effective in 
reasonably large systems, we used it to terminate chaos
in the two-dimensional Panfilov model, with $L = 500$
\cite{note2}.
Fig.~4 shows a sequence of images illustrating the evolution of 
chaos control when a sequence of biphasic rectangular pulses are applied
at the center. 
The time
necessary to achieve the controlled state, when the waves from the
stimulation point pervade the entire system, depends slightly on
the initial state of the system when the control is switched on.
Not surprisingly, we find that the stimulation frequency 
used to impose control in Fig.~4 belongs to a range for which
$f_{eff} > f_c$. 

Although most of the simulations were performed with the
Panfilov model , the arguments involving phase plane analysis
apply in general to excitable media having
a cubic-type nonlinearity. To ensure that our explanation 
is not sensitively model dependent we obtained
similar stimulation response diagrams for the Karma model \cite{Kar93}.

Some local control schemes 
envisage stimulating at special
locations, e.g., close to the tip of the spiral wave,
thereby driving the spiral wave towards the edges of
the system where they are absorbed \cite{Kri95}. 
However, aside from the fact that
spatiotemporal chaos involves a large number of coexisting spirals,
in a practical situation it may not be possible to have a choice regarding the
location of the stimulation point. We should therefore look for a
robust control method which is not critically sensitive to the position of 
the control point in the medium. 
There have been some proposals to use periodic
stimulation for controlling spatiotemporal chaos. 
For example, recently Zhang {\em et al}\cite{Zha03} have controlled 
some excitable media models by applying sinusoidal stimulation
at the center of the simulation domain. Looking
in detail into the mechanism of this type of control, we have come to
the conclusion that the key feature is the alternation between
positive and negative stimulation , i.e., biphasic pacing, and it is,
therefore, a special case of the general scheme presented here.

Previous explanations of why biphasic stimulation
is better than purely excitatory stimulation (that use only positive pulses),
have concentrated on the response to very large amplitude electrical shocks 
typically used in conventional defibrillation \cite{Kee99,And01} and 
have involved details of cardiac cell ion channels \cite{Jon00}.
To the best of our knowledge the present paper gives the simplest and 
most general picture
for understanding the efficacy of the biphasic scheme using 
very low amplitude perturbation, as it does not depend on the
details of ion channels responsible for cellular excitation.

There are some limitations to achieving control over a large spatial
domain in an excitable medium by pacing at
a particular point. Under some parameter regimes, the circular waves
propagating from this point may themselves become unstable and 
undergo conduction block at a distance from the origin
(similar to the process outlined in Ref.
\cite{Fox02}). In addition, the control requires a slightly higher
amplitude and has to be kept on for periods
much longer than spatially extended control methods \cite{Sin01}.
However, these drawbacks may be overcome if we use multiple stimulation
points arranged so that their regions of influence cover the entire
simulation domain. 

In conclusion, we have proposed a simple explanation of the efficacy
of low-amplitude 
biphasic stimulation in controlling spatiotemporal chaos in excitable
media (e.g., VF). It is based on the competition between the frequency
of the applied stimulation, and the effective frequency (obtained
from the dominant timescale) of chaos.
The former can be increased relative to the latter only by decreasing
the refractory period which is achieved by a negative stimulus
prior to applying the excitatory positive stimulation. Our analysis
makes it possible to design pacing waveforms for maximum efficiency
in controlling chaos.

\vspace{0.2cm}
We thank Sudeshna Sinha
for helpful comments. J.B. would like to thank DAAD for financial support.

\pagebreak

\begin{figure}[t!]
\centerline{\includegraphics[width=0.95\linewidth,clip] {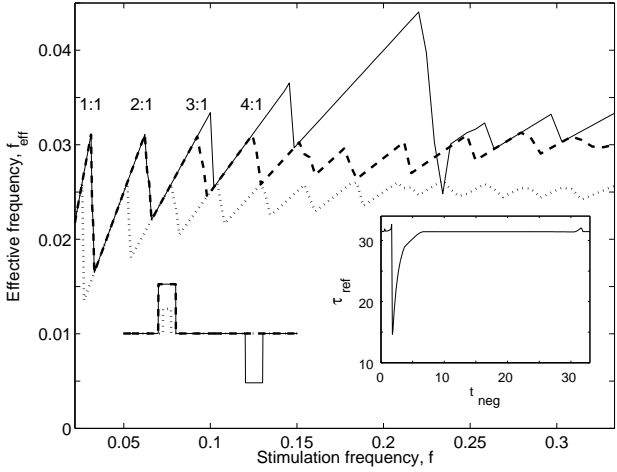}}
\caption{\small Stimulation response diagram for one-dimensional
Panfilov model ($L = 40$) for different stimulation frequencies $f$. 
The dotted and broken curves represent
purely excitatory
pulses of amplitude $A = 5$, pulse duration $\tau = 0.05 f^{-1}$, and $A = 10$,
$\tau = 0.1 f^{-1}$, respectively, while the
solid curve represents biphasic pulses of amplitude $A = 10$ and pulse duration
$\tau = 0.1 f^{-1}$ (as shown in the bottom left
corner). Note that, the highest effective 
frequencies $f_{eff}$ for the the three cases are very different.
The ratio $f : f_{eff}$ is shown for the first four peaks.
The inset shows the decrease in refractory period 
in absence of the diffusion term
when a negative stimulation
is applied at different times ($t_{neg}$) 
after the 
initial excitation.
}
\end{figure}


\begin{figure}[t!]
\centerline{\includegraphics[width=0.95\linewidth,clip] {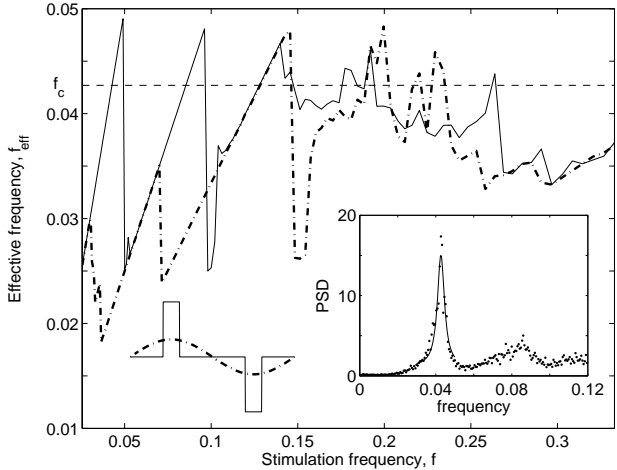}}
\caption{\small Stimulation response diagram for two-dimensional Panfilov 
model ($L = 26$) showing relative performance of different waveforms. 
The dash-dotted line represents a sinus wave ($A = 6$) and the solid curve
represents a wave of biphasic rectangular pulses ($A = 18.9$), as shown in
the bottom left corner, such that they have the same total energy.
The inset shows the power spectra of spatiotemporal chaos 
in the 2-D Panfilov model ($L = 500$). The $e$ variable was 
recorded for 3300 time units and the resulting power spectral density 
was averaged over 32 points. The peak occurs at the characteristic frequency
$f_c \simeq 0.0427$ which is indicated in the main figure by the broken line.
}
\end{figure}

\begin{figure}[t!]
\centerline{\includegraphics[width=0.95\linewidth,clip] {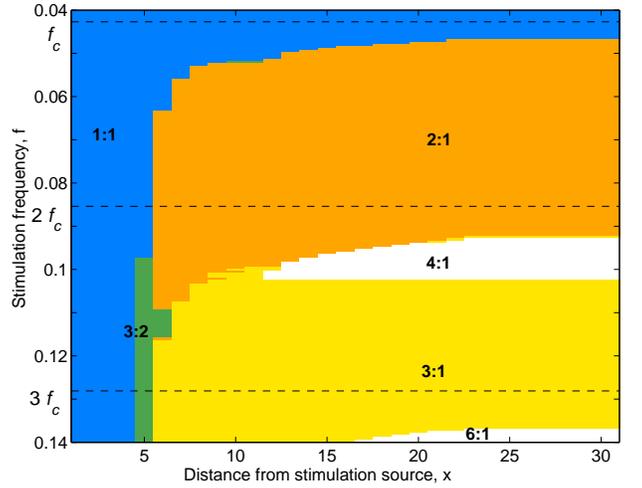}}
\vspace{0.5cm}
\caption{\small Distance dependence of stimulus response for different
stimulation frequencies $f$ in the two-dimensional
Panfilov model. Biphasic rectangular pulses ($A = 18.9$) having
duration $\tau = 0.1 f^{-1}$ are applied, which 
elicit a response having an effective frequency $f_{eff}$ at a particular
location. The first three cells ($x = 1,2,3$) are within the region
subject to direct stimulation.
The shaded regions represent 
different response ratios $f : f_{eff}$.
Integral multiples of the characteristic frequency $f_c$ are indicated
on the $f$-axis.
}
\end{figure}

\begin{figure}[t!]
\hspace{0.1cm}{\includegraphics[width=0.44\linewidth,clip] {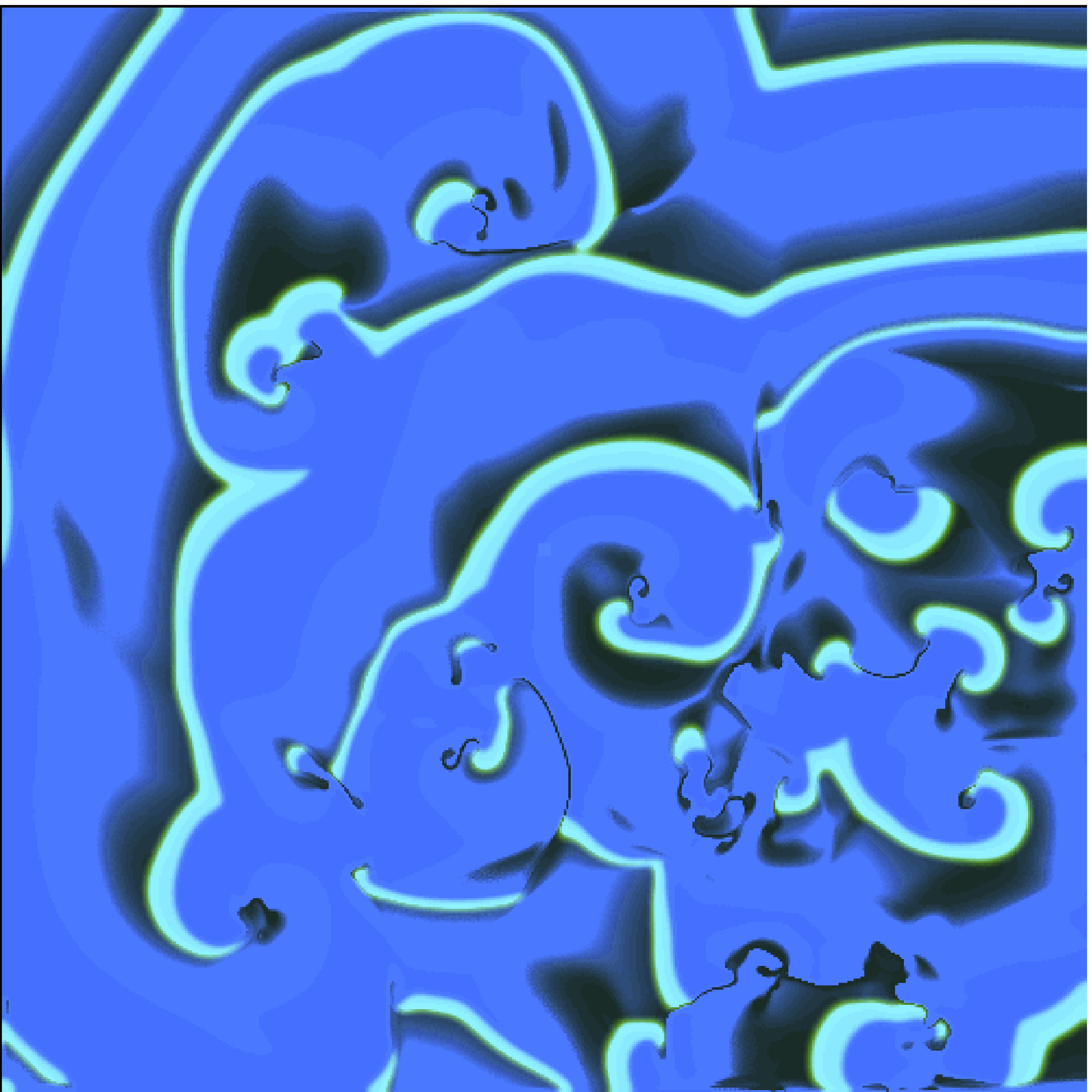}}
{\includegraphics[width=0.44\linewidth,clip] {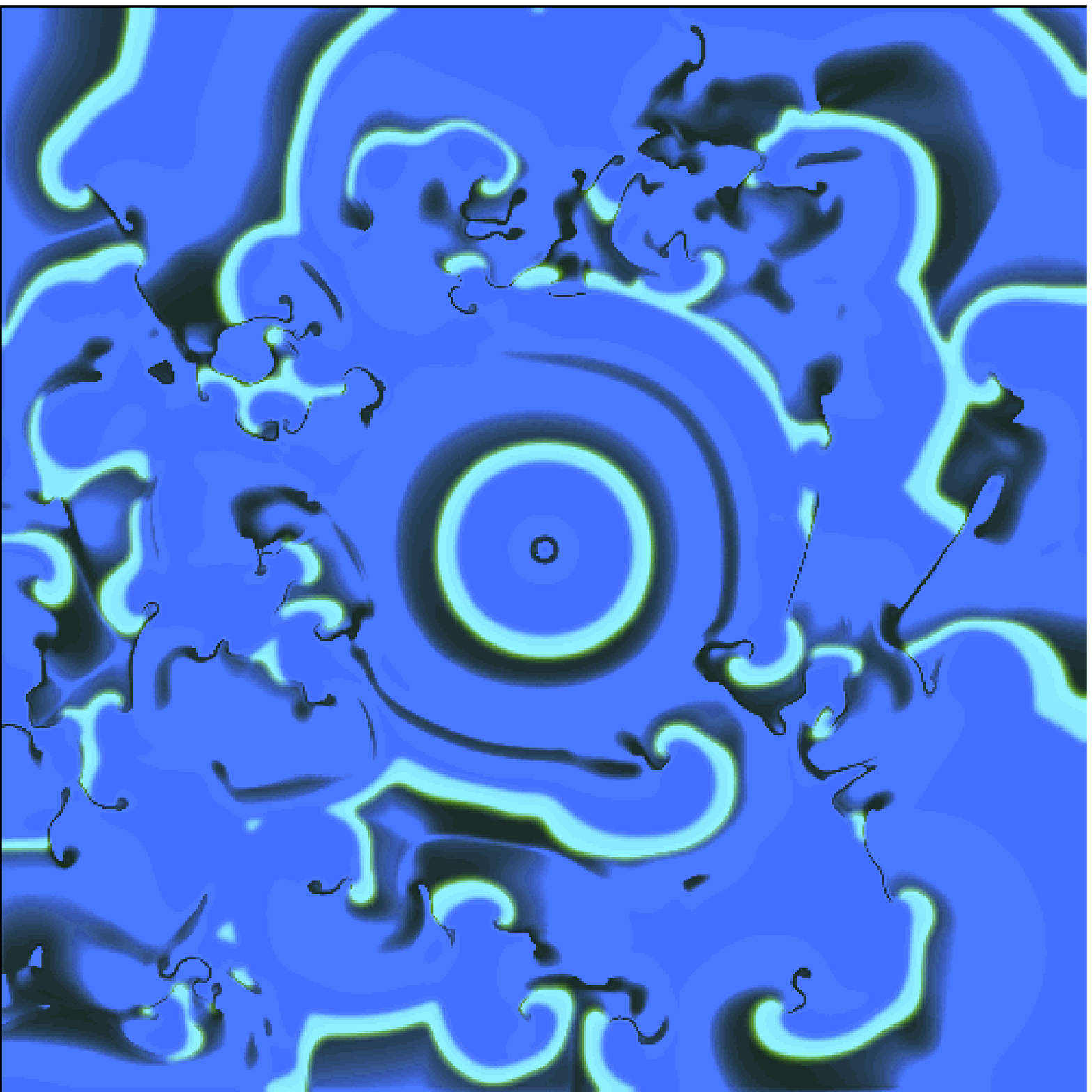}}

\vspace{0.1cm}
\hspace{0.1cm}{\includegraphics[width=0.44\linewidth,clip] {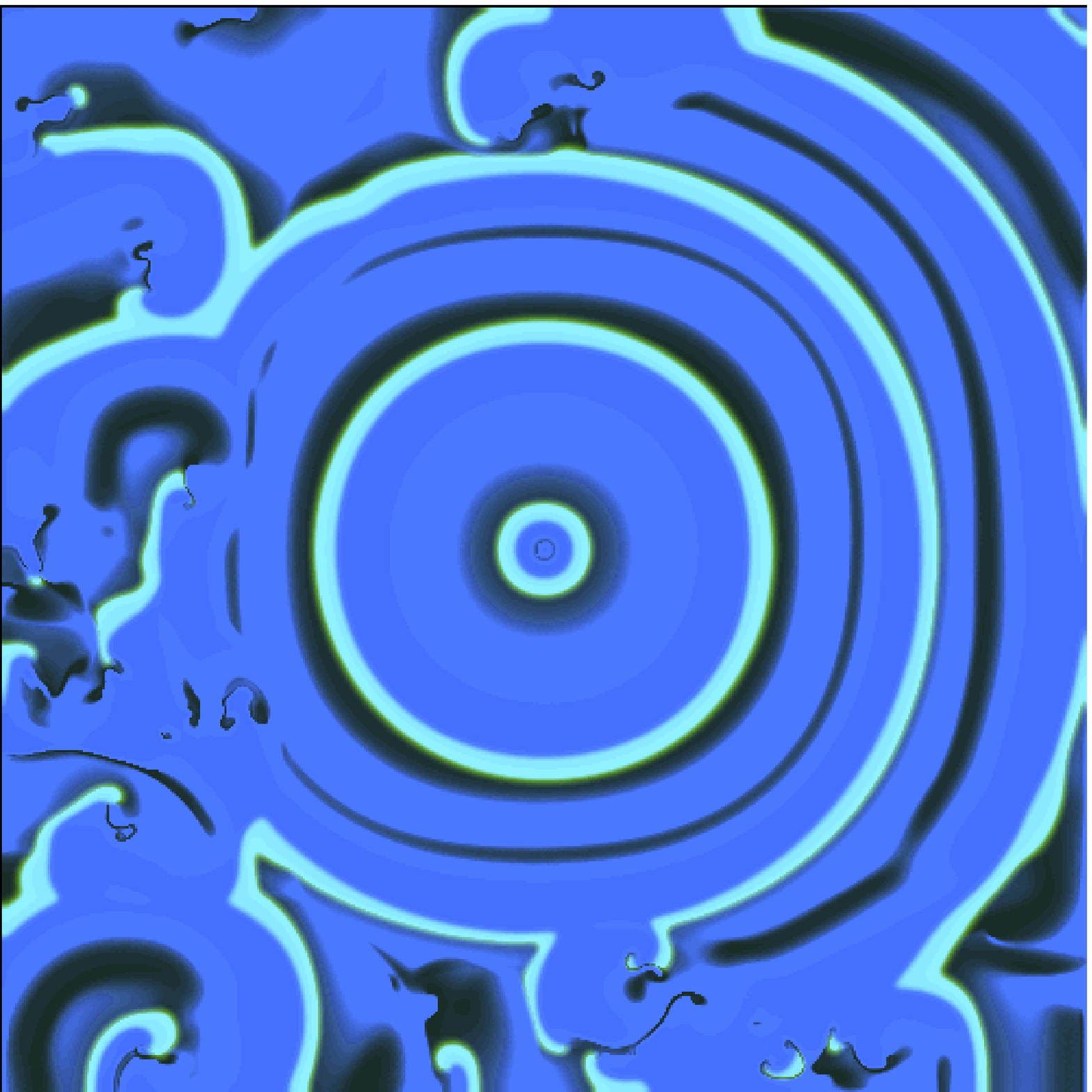}}
{\includegraphics[width=0.44\linewidth,clip] {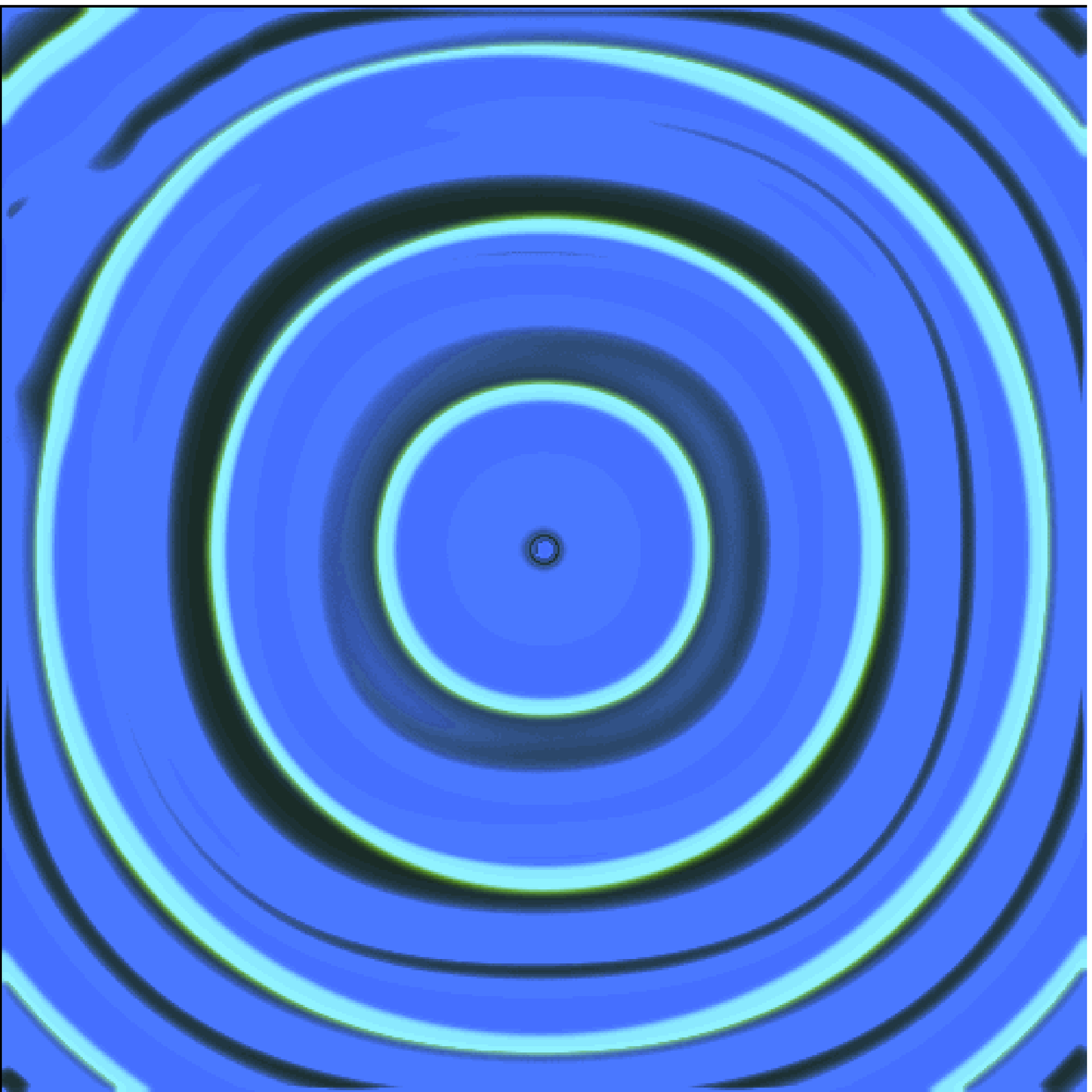}}
\vspace{0.5cm}
\caption{\small Control of spatiotemporal chaos in the two-
dimensional
Panfilov model ($L = 500$)
by applying biphasic pulses with amplitude $A$ = 18.9 and 
frequency $f = 0.13$ at the center of the simulation domain.
The pulse shape is rectangular, having a duration of $\sim 0.77 $ time units. 
Snapshots are shown for (top left) $t = 0$, 
(top right) $t = 1000$, (bottom left) $t = 2700$ and 
(bottom right) $t = 3800$ time units. The excitation wavefronts are
shown in white, black marks the recovered regions ready to be excited, 
while the shaded regions indicate different stages of refractoriness.}
\end{figure}


\begin{thebibliography}{10}
\bibitem{note1}
 The profile of the stimulated wave changes as it propagates along the
 medium, from biphasic at the stimulation source to gradually becoming 
 indistinguishable from
 a purely excitatory stimulation. As a result, far away
 from the source of stimulation, the response cannot have 
 a frequency higher than $\sim \tau_{ref}^{-1}$.
\bibitem{note2}
 The initial condition used for this purpose is a broken plane wave which
 is allowed to evolve for 5000 time units into a state displaying 
 spiral turbulence.
\end{thebibliography}
\end{document}